\documentclass[12pt,twoside]{article}
\usepackage{fleqn,espcrc1}

\usepackage{feynmp}                       

\usepackage[dvips]{color}                 
\usepackage{graphicx}                     
\usepackage{amssymb}                      

\usepackage{xspace}                       

\newcommand{\ia}{{\"{\i}}}   
\newcommand{\absatz}{\vspace{2ex}\noindent}

\newcommand{\journal}[4]{{#1}\textbf{{#2}}, #3 (#4)}

\newcommand{\NPA}{Nucl.\ Phys.\ \textbf{A}}
\newcommand{\NPB}{Nucl.\ Phys.\ \textbf{B}}
\newcommand{\PLB}{Phys.\ Lett.\ \textbf{B}}
\newcommand{\PRC}{Phys.\ Rev.\ \textbf{C}}

\newcommand{\dis}{\displaystyle}

\newcommand{\half}{\frac{1}{2}}

\newcommand{\ii}{\mathrm{i}}
\newcommand{\dd}{\mathrm{d}}
\newcommand{\tr}{\mathrm{tr}}
\newcommand{\T}{\mathrm{T}}

\newcommand{\pv}{\vec{\,\!p}\!\:{}}

\newcommand{\mpi}{m_\pi}
\newcommand{\fpi}{f_\pi}
\newcommand{\MeV}{\mathrm{MeV}}
\newcommand{\fm}{\mathrm{fm}}
\newcommand{\ENTNoPion}{ENT(${\pi\hskip-0.55em /}$)\xspace}
\newcommand{\NoPion}{{\pi\hskip-0.55em /}\xspace}
\newcommand{\upNoPion}{{}^{\pi\hskip-0.4em /}}

\newcommand{\de}{\partial}
\newcommand{\dev}{\vec{\de}}

\newcommand{\calA}{\mathcal{A}}
\newcommand{\calL}{\mathcal{L}}

\newcommand{\blue}{}

\begin{document}

\begin{fmffile}{fewfeyn}
  \fmfset{curly_len}{2mm} \fmfset{dash_len}{1.5mm} \fmfset{wiggly_len}{3mm}
  \newcommand{\feynbox}[2]{\mbox{\parbox{#1}{#2}}}
  \newcommand{\fs}{\scriptstyle}
  \newcommand{\hq}{\hspace{0.5em}} \newcommand{\hqm}{\hspace{-0.25em}}
  
  \fmfcmd{vardef ellipseraw (expr p, ang) = save radx; numeric radx; radx=6/10
    length p; save rady; numeric rady; rady=3/10 length p; pair center;
    center:=point 1/2 length(p) of p; save t; transform t; t:=identity xscaled
    (2*radx*h) yscaled (2*rady*h) rotated (ang + angle direction length(p)/2
    of p) shifted center; fullcircle transformed t enddef;
    style_def ellipse expr p= shadedraw ellipseraw (p,0); enddef; }

    
  \title{Effective Field Theory with Two and Three Nucleons\thanks{Invited
      plenary talk at the XVIIth European Conference on Few Body Problems in
      Physics, \'Evora (Portugal) 11th -- 16th September 2000; to be published
      in the Proceedings; preprint numbers nucl-th/0009058, TUM-T39-00-15.}  }
  
  \author{Harald W.~Grie\3hammer\address{
    Institut f{\"u}r Theoretische Physik, Physik-Department der\\
    Technischen Universit{\"a}t M{\"u}nchen, 85748 Garching, Germany\\
    Email: hgrie@physik.tu-muenchen.de}}
  
  \maketitle
  
\begin{abstract}
  Progress in the Effective Field Theory of two and three nucleon systems is
  sketched, concentrating on the low energy version in which pions are
  integrated out as explicit degrees of freedom. Examples given are
  calculations of deuteron Compton scattering, three body forces and the
  triton, and $\mathrm{nd}$ partial waves.
\end{abstract}

%

\section{Foundations of Effective Nuclear Theory}

This presentation is a cartoon of the Effective Field Theory (EFT) of two and
three nucleon systems as it emerged in the last three years, using a lot of
words and figures, and a few cheats. For details, I refer to the
bibliographies in~\cite{INTWorkshopSummary} as well as a recent
review~\cite{review}, and papers with J.-W.~Chen, R.P.~Springer and
M.J.~Savage~\cite{Compton}, P.F.~Bedaque~\cite{pbhg,pbfghg}, and
F.~Gabbiani~\cite{pbfghg}. I will mainly concentrate on the theory in which
pions are integrated out as explicit degrees of freedom, but also comment on
the extensions to include pions. B.~Holstein's talk~\cite{Holstein} provides
an introduction to EFT, and E.~Epelbaum's contribution~\cite{Epelbaum}
investigates one of the two proposals to include pions in more detail.

Effective Field Theory methods are largely used in many branches of physics
where a separation of scales exists~\cite{Holstein}. In low energy nuclear
systems, the scales are, on one side, the low scales of the typical momentum
of the process considered and the pion mass, and on the other side the higher
scales associated with chiral symmetry and confinement. This separation of
scales produces a low energy expansion, resulting in a description of strongly
interacting particles which is systematic and rigorous. It is also model
independent (meaning, independent of assumptions about the non-perturbative
QCD dynamics): Given that QCD is the theory of strong interactions and that
chiral symmetry is broken via the Goldstone mechanism, Wilson's
renormalisation group arguments show that there is only one local low energy
field theory which originates from it: Chiral Perturbation Theory and its
extension to the many-nucleon system discussed here.

\subsection{The Lagrangean}

Three main ingredients enter the construction of an EFT: The Lagrangean, the
power counting and a regularisation scheme.  First, the relevant degrees of
freedom have to be identified. In his original suggestion how to extend EFT
methods to systems containing two or more nucleons, Weinberg~\cite{Weinberg}
noticed that below the $\Delta$ production scale, only nucleons and pions need
to be retained as the infrared relevant degrees of freedom of low energy QCD.
Because at these scales the momenta of the nucleons are small compared to
their rest mass, the theory becomes non-relativistic at leading order in the
velocity expansion, with relativistic corrections systematically included at
higher orders. The most general chirally invariant Lagrangean consists hence
of contact interactions between non-relativistic nucleons, and between
nucleons and pions, with the first terms of the form
\begin{eqnarray}\label{ksw}
   {\mathcal{L}}_\mathrm{NN}&=&N^{\dagger}(\ii \de_0+\frac{\dev^2}{2M})N+
   \;\frac{\fpi^2}{8}\;
   \tr[(\partial_{\mu} \Sigma^{\dagger})( \partial^{\mu} \Sigma)]\;+
   \;g_A N^{\dagger} \vec{A}\cdot\sigma N\;-
   \\
   &&-\;C_0 (N^{\T} P^i N)^{\dagger} \ (N^{\T} P^i N)\;
   + \;\frac{C_2}{8}
   \left[(N^{\T} P^i N)^{\dagger} (N^{\T} P^i
     (\stackrel{\scriptscriptstyle\rightarrow}{\de}-
      \stackrel{\scriptscriptstyle\leftarrow}{\de})^2 N)+
   \mathrm{H.c.}\right]
   + \dots,\nonumber
\end{eqnarray}
where $N={p\choose n}$ is the nucleon doublet of two-component spinors and
$P^i$ is the projector onto the iso-scalar-vector channel, $
P^{i,\,b\beta}_{a\alpha}=\frac{1}{\sqrt{8}}
(\sigma_2\sigma^i)_{\alpha}^{\beta} (\tau_2)_a^b$.  The iso-vector-scalar part
of the $\mathrm{NN}$ Lagrangean introduces more constants $C_i$ and
interactions and has not been displayed for convenience. The field
$\xi(x)=\sqrt{\Sigma}=e^{\ii \Pi/\fpi}$ describes the pion.
$D_{\mu}=\partial_{\mu}+V_{\mu}$ is the chirally covariant derivative, and
$A_{\mu}=\frac{\ii}{2}(\xi\partial_{\mu}\xi^{\dagger}-\xi^{\dagger}
\partial_{\mu}\xi)$ ($V_\mu$) the axial (vector) pionic current. The
interactions involving pions are severely restricted by chiral invariance. As
such, the theory is an extension of Chiral Perturbation Theory and Heavy
Baryon Chiral Perturbation Theory to the many nucleon system. Like in its
cousins, the coefficients of the low energy Lagrangean encode all short
distance physics -- branes and strings, quarks and gluons, resonances like the
$\Delta$ or $\rho$ -- as strengths of the point-like interactions between
particles. As it is not possible yet to derive these constants by solving QCD
e.g.~on the lattice, the most practical way to determine them is by fitting to
experiment.

\subsection{The Power Counting}

Because the Lagrangean (\ref{ksw}) consists of infinitely many terms
only restricted by symmetry, an EFT may at first sight suffer from lack of
predictive power. Indeed, as the second part of an EFT formulation, predictive
power is ensured by establishing a power counting scheme, i.e.\ a way to
determine at which order in a momentum expansion different contributions will
appear, and keeping only and all the terms up to a given order. The
dimensionless, small parameter on which the expansion is based is the typical
momentum $Q$ of the process in units of the scale $\Lambda_\mathrm{NN}$ at
which the theory is expected to break down.  Values for $\Lambda_\mathrm{NN}$
and $Q$ have to be determined from comparison to experiments and are a priori
unknown. Assuming that all contributions are of natural size, i.e.\ ordered by
powers of $Q$, the systematic power counting ensures that the sum of all terms
left out when calculating to a certain order in $Q$ is smaller than the last
order retained, allowing for an error estimate of the final result.

For extremely small momenta, the pion does not enter as explicit degree of
freedom, and all its effects are absorbed into the coefficients $C_i$.  This
Effective Nuclear Theory with pions integrated out (\ENTNoPion) was recently
pushed to very high orders in the two-nucleon sector~\cite{CRS} where
accuracies of the order of $1\%$ were obtained. It can be viewed as a
systematisation of Effective Range Theory with the inclusion of relativistic
and short distance effects traditionally left out in that approach. Because of
non-analytic contributions from the pion cut, the breakdown scale of this
theory must be of the order $\upNoPion\Lambda_\mathrm{NN}\sim\mpi$. For the
ENT with explicit pions, we would suspect the breakdown scale to be of the
order of $M_\Delta-M$, since the $\Delta$ is not an explicit degree of freedom
in (\ref{ksw}).

Even if calculations of nuclear properties were possible starting from the
underlying QCD Lagrangean, EFT simplifies the problem considerably by
factorising it into a long distance part which contains the infrared-relevant
physics and is dealt with by EFT methods and a short distance part, subsumed
into the coefficients of the Lagrangean. QCD therefore ``only'' has to provide
these constants, avoiding full-scale calculations of e.g.~bound state
properties of two nuclear systems using quarks and gluons. EFT provides an
answer of finite accuracy because higher order corrections are systematically
calculable and suppressed in powers of $Q$. Hence, the power counting allows
for an error estimate of the final result, with the natural size of all
neglected terms known to be of higher order. Relativistic effects, chiral
dynamics and external currents are included systematically, and extensions to
include e.g.~parity violating effects are straightforward.  Gauged
interactions and exchange currents are unambiguous.  Results obtained with EFT
are easily dissected for the relative importance of the various terms.
Because only $S$-matrix elements between on-shell states are observables,
ambiguities nesting in ``off-shell effects'' are absent\footnote{Here, an
  extended note is in order. It has been noticed twenty years ago that an
  experiment cannot measure off-shell effects because in a field theory, no
  ambiguities arise from how we choose to define the interpolating
  fields~\cite{PolitzerArzt}.  We can consistently neglect all operators that
  vanish by the equations of motion because each of them makes a contribution
  to an observable that has exactly the same form as higher dimension
  operators that are present in the theory. Further, such operators can be
  removed by field redefinitions, and therefore it is consistent to work with
  a Lagrange density that does not contain operators that vanish by the
  equations of motion. Explicit examples can be found e.g.~in \cite{pbhg} and
  \cite[App.~B]{KSW2}. Of course, particles in loops still propagate off-shell
  in an EFT, just as usual.}. On the other hand, because only symmetry
considerations enter the construction of the Lagrangean, EFTs are less
restrictive as no assumption about the underlying QCD dynamics is
incorporated. Hence the proverbial quib that ``EFT parameterises our
ignorance''.

\absatz In systems involving two or more nucleons, establishing a power
counting is complicated because unnaturally large scales have to be
accommodated: Given that the typical low energy scale in the problem should be
the mass of the pion as the lightest particle emerging from QCD, fine tuning
is required to produce the large scattering lengths in the $\mathrm{S}$ wave
channels ($1/a^{{}^1\mathrm{S}_0}=-8.3\;\MeV,\;
1/a^{{}^3\mathrm{S}_1}=36\;\MeV$).  Since there is a bound state in the
${}^3\mathrm{S}_1$ channel with a binding energy $B=2.225\;\MeV$ and hence a
typical binding momentum $\gamma=\sqrt{M B}\simeq 46\;\MeV$ well below the
scale $\Lambda_\mathrm{NN}$ at which the theory should break down, it is also
clear that at least some processes have to be treated non-perturbatively in
order to accommodate the deuteron, i.e.~an infinite number of diagrams has to
be summed.

For simplicity, let us first turn to the EFT in which the pion is integrated
out into the coefficients $\upNoPion C_i$ of the Lagrangean. Here, a way to
incorporate this fine tuning into the power counting was suggested by Kaplan,
Savage and Wise~\cite{KSW}, and by van Kolck~\cite{BiraAleph}. At very low
momenta, contact interactions with several derivatives -- like $p^2\,\upNoPion
C_2$ -- should become unimportant, and we are left only with the contact
interactions proportional to $\upNoPion C_0$.  The leading order contribution
to nucleons scattering in an $\mathrm{S}$ wave comes hence from four nucleon
contact interactions and is summed geometrically as in
Fig.~\ref{fig:deuteronprop} in order to produce the shallow real bound state,
i.e.~the deuteron.
\begin{figure}[!htb]
  \begin{center}
    $\setlength{\unitlength}{1.3pt}
        \feynbox{25\unitlength}{
            \begin{fmfgraph*}(25,25)
              \fmfleft{i} \fmfright{o} \fmf{double,width=thin}{i,o}
            \end{fmfgraph*}}
          \hq=\hq\feynbox{30\unitlength}{
            \begin{fmfgraph*}(30,25)
              \fmfleft{i1,i2} \fmfright{o1,o2}
              \fmf{vanilla,width=thin}{i1,v,o2}
              \fmf{vanilla,width=thin}{i2,v,o1}
            \end{fmfgraph*}} +
          \feynbox{45\unitlength}{
            \begin{fmfgraph*}(45,25)
              \fmfleft{i1,i2} \fmfright{o1,o2}
              \fmf{vanilla,width=thin,tension=6}{i1,v1}
              \fmf{vanilla,width=thin,tension=6}{v2,o2}
              \fmf{vanilla,width=thin,tension=6}{i2,v1}
              \fmf{vanilla,width=thin,tension=6}{v2,o1}
              \fmf{vanilla,width=thin,left=0.5}{v1,v2}
              \fmf{vanilla,width=thin,left=0.5}{v2,v1}
            \end{fmfgraph*}}  +
          \feynbox{63\unitlength}{
            \begin{fmfgraph*}(63,25)
              \fmfleft{i1,i2} \fmfright{o1,o2}
              \fmf{vanilla,width=thin,tension=6}{i1,v1}
              \fmf{vanilla,width=thin,tension=6}{v2,o2}
              \fmf{vanilla,width=thin,tension=6}{i2,v1}
              \fmf{vanilla,width=thin,tension=6}{v2,o1}
              \fmf{vanilla,width=thin,left=0.7}{v1,v3}
              \fmf{vanilla,width=thin,left=0.7}{v3,v1}
              \fmf{vanilla,width=thin,left=0.7}{v2,v3}
              \fmf{vanilla,width=thin,left=0.7}{v3,v2}
            \end{fmfgraph*}}  +\;\dots
     \;=\;\dis\frac{-\upNoPion C_0}{1-\rule{0pt}{16pt}\feynbox{24\unitlength}{
            \begin{fmfgraph*}(25,12)
              \fmfleft{i} \fmfright{o} \fmf{vanilla,width=thin,left=0.5}{i,o}
              \fmf{vanilla,width=thin,left=0.5}{o,i}
            \end{fmfgraph*}}}
        $
\vspace{-15pt}
\setlength{\unitlength}{1pt}
    \caption{Re-summation of the contact interactions into the deuteron
      propagator.}
    \label{fig:deuteronprop}
  \end{center}
\vspace*{-4ex}
\end{figure}

How to justify this? Dimensional analysis allows any diagram to be estimated
by scaling momenta by a factor of $Q$ and non-relativistic kinetic energies by
a factor of $Q^2/M$.  The remaining integral includes no dimensions and is
taken to be of the order $Q^0$ and of natural size. This scaling implies the
rule that nucleon propagators contribute one power of $M/Q^2$ and each loop a
power of $Q^5/M$.  Assuming that
\begin{eqnarray}\label{scalingksw}
  \upNoPion C_0\sim\frac{1}{M Q}&\;\;,\;\;&
  \upNoPion C_2\sim\frac{1}{M \;\upNoPion\Lambda_\mathrm{NN} Q^2}\;\;,
\end{eqnarray}
the diagrams contributing at leading order to the deuteron propagator are
indeed an infinite number as shown in Fig.~\ref{fig:deuteronprop}, each one of
the order $1/(MQ)$. The deuteron propagator
\begin{equation}
  \label{deuteronpropagator}
  \frac{4\pi}{M}\;\frac{-\ii}{\frac{4\pi}{M\,\upNoPion C_0}+\mu-
    \sqrt{\frac{\pv^2}{4}-Mp_0-\ii\varepsilon}}
\end{equation}
has the correct pole position and cut structure when one chooses
$
\upNoPion C_0(\mu)=\frac{4\pi}{M}\;\frac{1}{\gamma-\mu}
$.

$\upNoPion C_0$ becomes dependent on an arbitrary scale $\mu$ because of the
regulator dependent, linear UV divergence in each of the bubble diagrams.
Indeed, when choosing $\mu\sim Q$, the leading order contact interaction
scales as demanded in (\ref{scalingksw}). As expected for a physical
observable, the $\mathrm{NN}$ scattering amplitude becomes independent of
$\mu$, the renormalisation scale or cut-off chosen.  All other coefficients
$\upNoPion C_i$ can be shown to be higher order, so that the scheme is
self-consistent. Observables are independent of the cut-off chosen.

The linear divergence does not show in dimensional regularisation as a pole in
$4$ dimensions, but it does appear as a pole in $3$ dimensions which we
subtract following the Power Divergence Subtraction scheme~\cite{KSW}.
Dimensional regularisation is chosen to explicitly preserve the systematic
power counting as well as all symmetries (esp.~chiral invariance) at each
order in every step of the calculation. At leading (LO), next-to-leading order
(NLO) and often even N3LO in the two nucleon system, it also allows for
simple, closed answers whose analytic structure is readily asserted. Power
Divergence Subtraction moves hence a somewhat arbitrary amount of the short
distance contributions from loops to counterterms and makes precise
cancellations manifest which arise from fine tuning~\cite{BiraAleph}.

\subsection{Including Pions}

The power counting of the zero and one nucleon sector of the theory are fixed
by Chiral Perturbation Theory and its extension to the one baryon sector.
Therefore, the question to be posed is: How does the power counting of the
contact terms $\upNoPion C_i$ change above the pion cut, i.e.~when the pion
must be included as explicit degree of freedom? We do not know yet, but the
following two promising proposals are on the market.

\begin{figure}[!ht]
  \begin{center}
    \vspace{-2ex}
    \feynbox{90\unitlength}{
            \begin{fmfgraph*}(90,45)
              \fmfleft{i} \fmfright{o} \fmf{double,width=thin,tension=8}{i,v1}
              \fmf{double,width=thin,tension=8}{v2,o}
              \fmf{vanilla,width=thin,left=0.65}{v1,v2}
              \fmf{vanilla,width=thin,left=0.65}{v2,v1} \fmffreeze
              \fmffreeze \fmfipath{pa} \fmfiset{pa}{vpath(__v1,__v2)}
              \fmfipath{pb} \fmfiset{pb}{vpath(__v2,__v1)} \fmfi{dashes}{point
                1/2 length(pa) of pa -- point 1/2 length(pb) of pb}
              \end{fmfgraph*}}\hq+\hq
            \feynbox{135\unitlength}{
            \begin{fmfgraph*}(135,60)
              \fmfleft{i} \fmfright{o} \fmf{double,width=thin,tension=5}{i,v1}
              \fmf{double,width=thin,tension=5}{v2,o}
              \fmf{vanilla,width=thin,left=0.8}{v1,v3}
              \fmf{vanilla,width=thin,left=0.8}{v3,v1}
              \fmf{vanilla,width=thin,left=0.8}{v2,v3}
              \fmf{vanilla,width=thin,left=0.8}{v3,v2}
              \fmfv{decor.shape=square,decor.size=6,
                label=$p^2C_2$,
                label.angle=90,label.dist=0.18w}{v3}
            \end{fmfgraph*}}
\vspace{-20pt}
          \caption{The next-to-leading order in ENT(KSW).}
    \label{fig:NLOKSW}
  \end{center}
\vspace*{-4ex}
\end{figure}
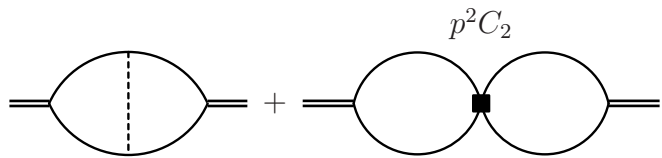

If pulling out pion effects does not affect the running of $C_0$ too much, one
surprising result arises: Because chiral symmetry implies a derivative
coupling of the pion to the nucleon at leading order, the instantaneous one
pion exchange scales as $Q^0$ and is {\it smaller} than the contact piece
${}^\mathrm{KSW}C_0\sim \upNoPion C_0 \sim Q^{-1}$. Pion exchange and higher
derivative contact terms appear hence only as perturbations at higher orders.
The LO contribution in this scheme is still given by the geometric series in
Fig.~\ref{fig:deuteronprop}.  In contradistinction to iterative potential
model approaches, each higher order contribution is inserted only once.  In
this scheme, the only non-perturbative physics responsible for nuclear binding
is extremely simple, and the more complicated pion contributions are at each
order given by a finite number of diagrams. For example, the NLO contributions
to the deuteron are the one instantaneous pion exchange and the four nucleon
interaction with two derivatives, Fig.~\ref{fig:NLOKSW}.  The constants are
determined e.g.~by demanding the correct deuteron pole position and residue.
This approach is known as the ``KSW'' counting of ENT, ENT(KSW)~\cite{KSW}.
\setlength{\unitlength}{1pt}
    
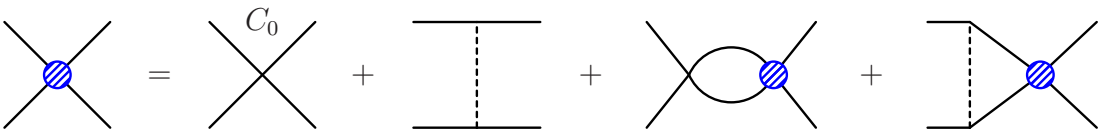
\begin{figure}[!ht]
  \begin{center}
    \vspace{-2ex}
    $
  \feynbox{50\unitlength}{
            \begin{fmfgraph*}(50,40)
              \fmfright{i1,i2}
              \fmfleft{o1,o2}
              \fmf{vanilla}{i1,v,o1}
              \fmf{vanilla}{i2,v,o2}
              \fmfv{decor.shape=circle,decor.filled=shaded,foreground=blue,
                decor.size=10}{v}
            \end{fmfgraph*}}
          \hq=\hq
  \feynbox{50\unitlength}{
            \begin{fmfgraph*}(50,40)
              \fmfright{i1,i2} \fmfleft{o1,o2} \fmf{vanilla}{i1,v,o1}
              \fmf{vanilla}{i2,v,o2}
              \fmfv{label=$C_0$,label.angle=90,
                label.dist=0.3w}{v}
            \end{fmfgraph*}}
          \hq+\hq
   \feynbox{60\unitlength}{
            \begin{fmfgraph*}(60,40)
              \fmfright{i1,i2}
              \fmfleft{o1,o2}
              \fmf{vanilla}{i1,v1,o1}
              \fmf{vanilla}{i2,v2,o2}
              \fmffreeze
              \fmf{dashes}{v1,v2}
            \end{fmfgraph*}}
          \hq+\hq
   \feynbox{80\unitlength}{
            \begin{fmfgraph*}(80,40)
              \fmfright{i1,i2}
              \fmfleft{o1,o2}
              \fmf{vanilla}{i1,v1}
              \fmf{vanilla}{i2,v1}
              \fmf{vanilla}{o1,v2}
              \fmf{vanilla}{o2,v2}
              \fmf{vanilla,left=0.65,tension=0.5}{v1,v2}
              \fmf{vanilla,left=0.65,tension=0.5}{v2,v1}
              \fmfv{decor.shape=circle,decor.filled=shaded,foreground=blue,
                decor.size=10}{v1}
            \end{fmfgraph*}}
          \hq+\hq
   \feynbox{80\unitlength}{
            \begin{fmfgraph*}(80,40)
              \fmfright{i1,i2}
              \fmfleft{o1,o2}
              \fmf{vanilla}{i1,v1}
              \fmf{vanilla}{i2,v1}
              \fmf{phantom}{i1,v2}
              \fmf{phantom,tension=3}{v2,o1}
              \fmf{phantom}{i2,v3}
              \fmf{phantom,tension=3}{v3,o2}
              \fmf{phantom}{i1,v1,o2}
              \fmf{phantom}{i2,v1,o1}
              \fmffreeze
              \fmf{vanilla}{v1,v2}
              \fmf{vanilla}{v2,o1}
              \fmf{vanilla}{v1,v3}
              \fmf{vanilla}{v3,o2}
              \fmf{dashes}{v2,v3}
              \fmfv{decor.shape=circle,decor.filled=shaded,foreground=blue,
                decor.size=10}{v1}
            \end{fmfgraph*}}
          $
\vspace{-10pt}
    \caption{The leading order in ENT(Weinberg).}
    \label{fig:weinberg}
  \end{center}
\vspace*{-4ex}
\end{figure}

If on the other hand, the power counting for $\upNoPion C_0$ is dramatically
modified when one includes pions, both the $C_0$ interactions and the one pion
exchange might have to be iterated. In his paper, Weinberg~\cite{Weinberg}
therefore suggested to power count not the amplitude -- as is usually done in
EFT -- but the potential, and then to solve a Schr\"odinger equation with a
chiral potential, as pictorially represented in Fig.~\ref{fig:weinberg}.
E.~Epelbaum's talk~\cite{Epelbaum} gives more details on the results obtained
so far in this approach.

Both power countings are presently under investigation for consistency and
convergence, and each has its advantages and shortcomings~\cite{review}. If
both approaches are self-consistent, Nature will tell us which it chose.
Although in general process dependent, the expansion parameter is found to be
of the order of $\frac{1}{3}$ in most applications, so that NLO calculations
can be expected to be accurate to about $10\%$, and N2LO calculations to about
$4\%$.  In all cases, experimental agreement is within the estimated
theoretical uncertainties, and in some cases, previously unknown counterterms
could be determined.

\section{Applications}

\subsection{Deuteron Compton Scattering}

The first example we turn to demonstrates not only how simple it is to compute
processes involving both external gauge and exchange currents in a
self-consistent way, but especially how powerful the power counting is in
estimates of theoretical uncertainties.

\begin{figure}[!htb]
  \begin{center}
    \vspace{-2ex}
    \includegraphics*[width=0.49\textwidth]{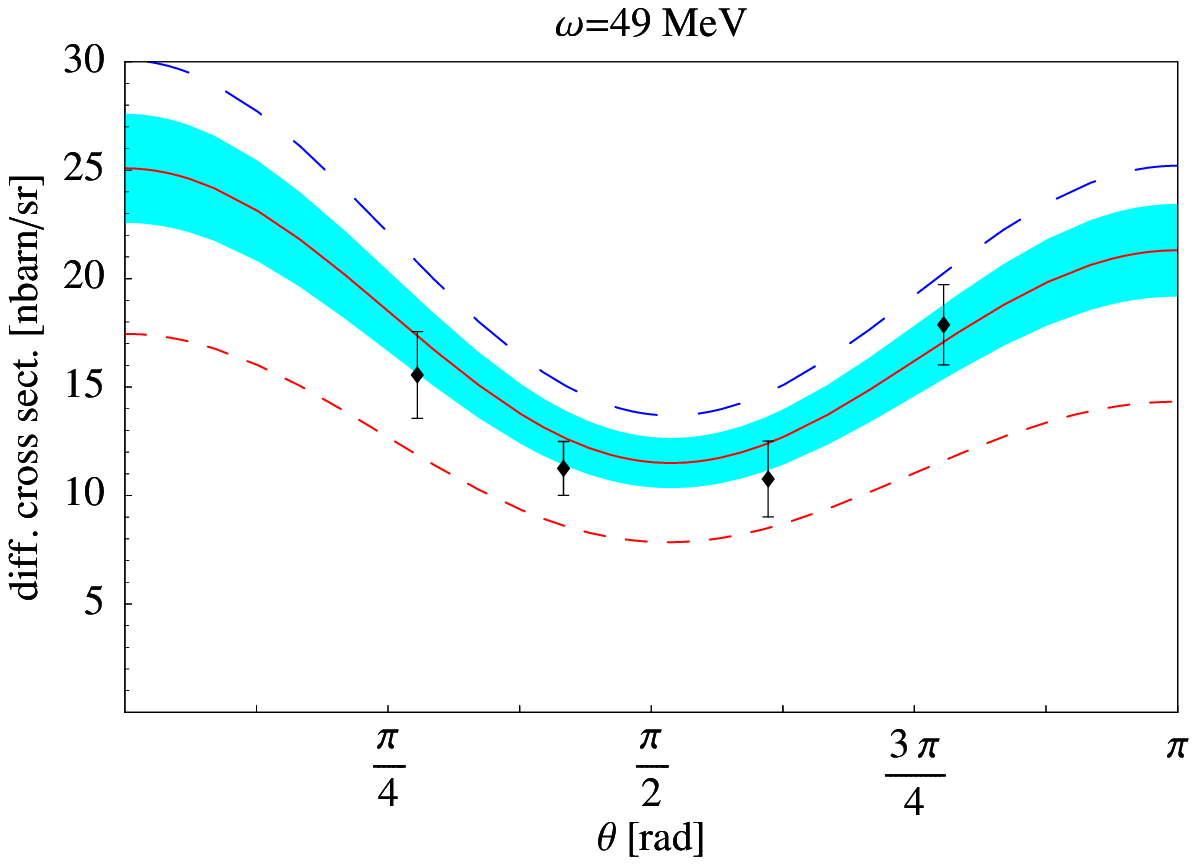} \hfill
    \includegraphics*[width=0.49\textwidth]{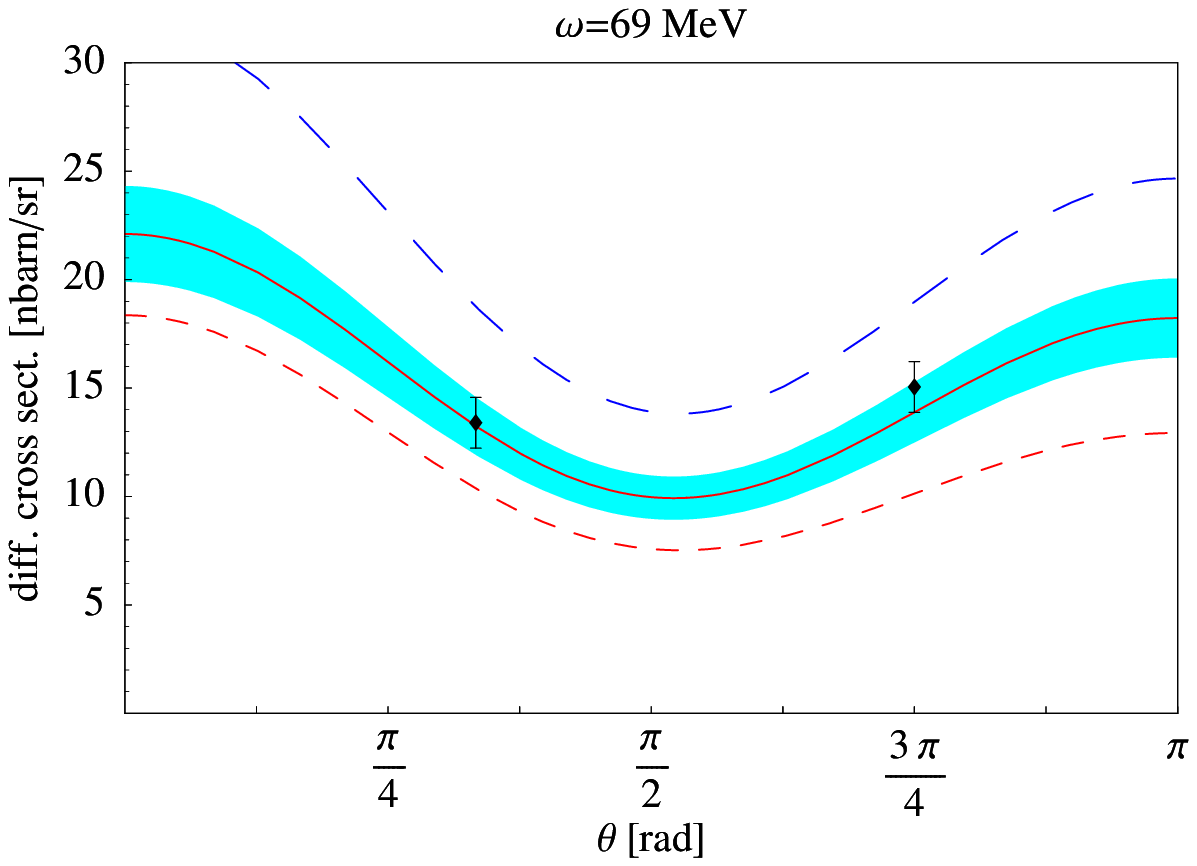}
  \vspace{-15pt}
\caption{The differential cross section for elastic 
  \protect$\gamma$-deuteron Compton scattering at incident photon energies of
  \protect$E_{\gamma}=49\ \mathrm{MeV}$ and \protect$69\ \mathrm{MeV}$ in
  EFT(KSW)\protect~\cite{Compton}, no free parameters. Dashed: LO;
  long dashed: 
  NLO without the graphs that contribute to the nucleon polarisability; solid
  curve: complete NLO result. The estimate of the accuracy of the calculation
  at NLO (\protect$\pm 10\%$) is indicated by the shaded area.}
\label{fig:compton}
  \end{center} 
\vspace*{-4ex}
\end{figure}

Using the power counting ENT(KSW) in which pions are perturbative, the elastic
deuteron Compton scattering cross section~\cite{Compton} to NLO is
parameter-free with an accuracy of $10\%$.  Contributions at NLO include the
pion graphs that dominate the electric polarisability of the nucleon
$\alpha_E$ from their $\frac{1}{m_\pi}$ behaviour in the chiral limit. Because
they are NLO, the power counting predicts their contribution to be sizeable,
namely $20$ to $30\%$. The comparison with experiment in
Fig.~\ref{fig:compton} shows good agreement and therefore confirms the
HB$\chi$PT value for $\alpha_E$. The deuteron scalar and tensor electric and
magnetic polarisabilities are also easily extracted~\cite{Compton}.

Compton scattering has also been investigated in
ENT(Weinberg)~\cite{Beaneetal} and in \ENTNoPion\ \cite{hggrms}. Surprisingly,
the results all agree within the theoretical uncertainty of the calculations
even at relatively high momenta. This seems to suggest that non-analytic pion
contributions from meson exchange diagrams are small.

\subsection{Three Body Forces and the Triton}

Since all interactions permitted by the symmetries must be included into the
Lagrangean, EFT dictates that in the three body sector, interactions like
\begin{equation}
  \label{threebodyterm}
  \calL_\mathrm{3body}=H\;(N^\dagger N)^3
\end{equation}
with unknown strength $H$ are present. Turning again to \ENTNoPion, we
therefore have to ask at which order in the power counting they will start to
contribute. Bedaque, Hammer and van Kolck~\cite{Stooges2} found the surprising
result that an unusual renormalisation makes the three body force of leading
order in the triton channel.

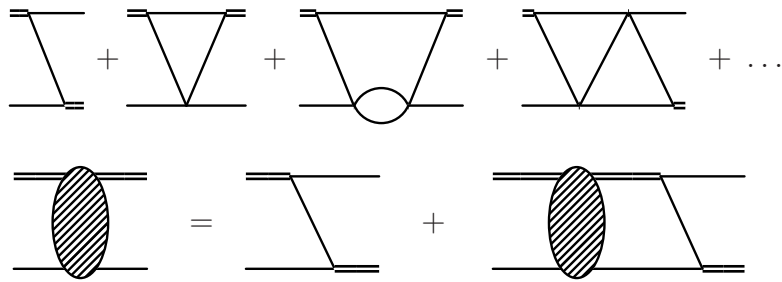
\begin{figure}[!htb]
  \begin{center}
    \vspace{-2ex}
    \setlength{\unitlength}{0.7pt}
    \feynbox{50\unitlength}{
            \begin{fmfgraph*}(50,50)
              \fmfleft{i2,i1} \fmfright{o2,o1}
              \fmf{double,tension=6}{i1,v1,v2} \fmf{vanilla,width=thin}{v2,o1}
              \fmf{double,tension=6}{v3,v4,o2} \fmf{vanilla,width=thin}{i2,v3}
              \fmffreeze \fmf{vanilla,width=thin}{v2,v3}
            \end{fmfgraph*}}
          \hqm$+$\hqm \feynbox{80\unitlength}{
            \begin{fmfgraph*}(80,50)
              \fmfleft{i2,i1} \fmfright{o2,o1} \fmf{double,tension=4}{i1,v1}
              \fmf{vanilla,width=thin}{v1,v2} \fmf{double,tension=4}{o1,v2}
              \fmf{vanilla,width=thin}{i2,v3} \fmf{vanilla,width=thin}{v3,o2}
              \fmffreeze \fmf{vanilla,width=thin}{v2,v3}
              \fmf{vanilla,width=thin}{v3,v1}
            \end{fmfgraph*}}
          \hqm$+$\hqm \feynbox{110\unitlength}{
            \begin{fmfgraph*}(110,50)
              \fmfleft{i2,i1} \fmfright{o2,o1} \fmf{double,tension=8}{i1,v1}
              \fmf{vanilla,width=thin}{v1,v2} \fmf{double,tension=8}{o1,v2}
              \fmf{double,tension=100}{v3,v3a}
              \fmf{double,tension=100}{v4,v4a} \fmf{phantom}{v3a,v4a}
              \fmf{vanilla,width=thin}{i2,v3} \fmf{vanilla,width=thin}{v4,o2}
              \fmffreeze \fmf{vanilla,width=thin}{v2,v4}
              \fmf{vanilla,width=thin}{v3,v1} \fmffreeze
              \fmf{vanilla,width=thin,left=0.65}{v3a,v4a}
              \fmf{vanilla,width=thin,left=0.65}{v4a,v3a}
            \end{fmfgraph*}}
          \hqm$+$\hqm \feynbox{110\unitlength}{
            \begin{fmfgraph*}(110,50)
              \fmfleft{i2,i1} \fmfright{o2,o1} \fmf{double,tension=8}{i1,v1}
              \fmf{vanilla,width=thin}{v1,v4} \fmf{double,tension=100}{v4,v5}
              \fmf{vanilla,width=thin,tension=1.666}{v5,o1}
              \fmf{double,tension=8}{o2,v6} \fmf{vanilla,width=thin}{v6,v3}
              \fmf{double,tension=100}{v3,v2}
              \fmf{vanilla,width=thin,tension=1.666}{v2,i2} \fmffreeze
              \fmf{vanilla,width=thin}{v1,v2} \fmf{vanilla,width=thin}{v3,v4}
              \fmf{vanilla,width=thin}{v5,v6}
            \end{fmfgraph*}}
          \hqm$+\;\dots$
          \\[5ex]
    \feynbox{90\unitlength}{
            \begin{fmfgraph*}(90,50)
              \fmfleft{i2,i1} \fmfright{o2,o1}
              \fmf{double,width=thin,tension=3}{i1,v1}
              \fmf{double,width=thin,tension=1.5}{v1,v3,v2}
              \fmf{double,width=thin,tension=3}{v2,o1}
              \fmf{vanilla,width=thin}{i2,v4,o2} \fmffreeze \fmffreeze
              \fmf{ellipse,rubout=1}{v3,v4}
              \end{fmfgraph*}}
            \hq$=$\hq \feynbox{90\unitlength}{
            \begin{fmfgraph*}(90,50)
              \fmfleft{i2,i1} \fmfright{o2,o1}
              \fmf{double,width=thin,tension=4}{i1,v1,v2}
              \fmf{vanilla,width=thin}{v2,o1}
              \fmf{double,width=thin,tension=4}{v3,v4,o2}
              \fmf{vanilla,width=thin}{i2,v3} \fmffreeze
              \fmf{vanilla,width=thin}{v2,v3}
              \end{fmfgraph*}}
            \hq$+$\hq \feynbox{170\unitlength}{
            \begin{fmfgraph*}(170,50)
              \fmfleft{i2,i1} \fmfright{o2,o1}
              \fmf{double,width=thin,tension=3}{i1,v1}
              \fmf{double,width=thin,tension=1.5}{v1,v6,v5}
              \fmf{double,width=thin,tension=3}{v5,v2}
              \fmf{vanilla,width=thin}{v2,o1} \fmf{vanilla,width=thin}{i2,v7}
              \fmf{vanilla,width=thin,tension=0.666}{v7,v4}
              \fmf{double,width=thin,tension=4}{v4,v3,o2} \fmffreeze
              \fmf{vanilla,width=thin}{v4,v2} \fmf{ellipse,rubout=1}{v6,v7}
              \end{fmfgraph*}}
    \vspace{-10pt}
    \caption{The double infinite series of LO
      ``pinball'' diagrams, some of which are shown in the first line, is
      equivalent to the solution of the Faddeev equation of the second line.}
    \label{fig:LOfaddeev}
  \end{center}
  \vspace*{-4ex}
\end{figure}

In order to understand this finding, let us first consider the LO diagrams
which come from two nucleon interactions. The absence of Coulomb interactions
in the $\mathrm{nd}$ system ensures that only properties of the strong
interactions are probed.  All graphs involving only $\upNoPion C_0$
interactions are of the same order and form a double series which cannot be
written down in closed form.  Summing all ``bubble-chain'' sub-graphs into the
deuteron propagator, one can however obtain the solution numerically from the
integral equation pictorially shown in Fig.~\ref{fig:LOfaddeev} within seconds
on a personal computer.

Because one nucleon is exchanged in the intermediate state, each diagram is of
the order of the nucleon propagator, i.e.~$Q^{-2}$, while the first three
buddy force (\ref{threebodyterm}) seems to be of order $Q^0$. We are therefore
tempted to assume that three body forces are at worst N2LO, i.e.~of the order
$10\%$. In the quartet channel, power counting suggests even N4LO because the
Pauli principle forbids three body forces without derivatives. However, such
na{\ia}ve counting was already fallacious in the two body sector because of
the presence of a low lying bound state and of a linear divergence in each of
the bubble graphs making up Fig.~\ref{fig:deuteronprop}. We therefore
investigate more carefully the UV behaviour of the amplitude $\calA(k,p)$ at
half off-shell momenta $p\gg k,\gamma$. The integral equation simplifies then
to
\begin{eqnarray}
  \label{UVfaddeev}
        \calA_{(l,s)}(0,p)&=&
        \frac{4\;\lambda(s)}{\sqrt{3}\,\pi}
          \int\limits_0^\infty \frac{\dd q}{p}
          \;\calA_{(l,s)}(0,q)\;
          Q_{l}\left[\frac{p}{q}+ \frac{q}{p}\right]\;\;,
\end{eqnarray}
where $\lambda(s)=-\frac{1}{2}$ in the spin quartet channel, and $1$ in
the doublet. $Q_l$ is the Legendre Polynomial of the second kind, $l$ the
angular momentum of the partial wave investigated.

\begin{figure}[!ht]
  \begin{center}
    \vspace{-2ex}
    \includegraphics*[width=0.47\linewidth]{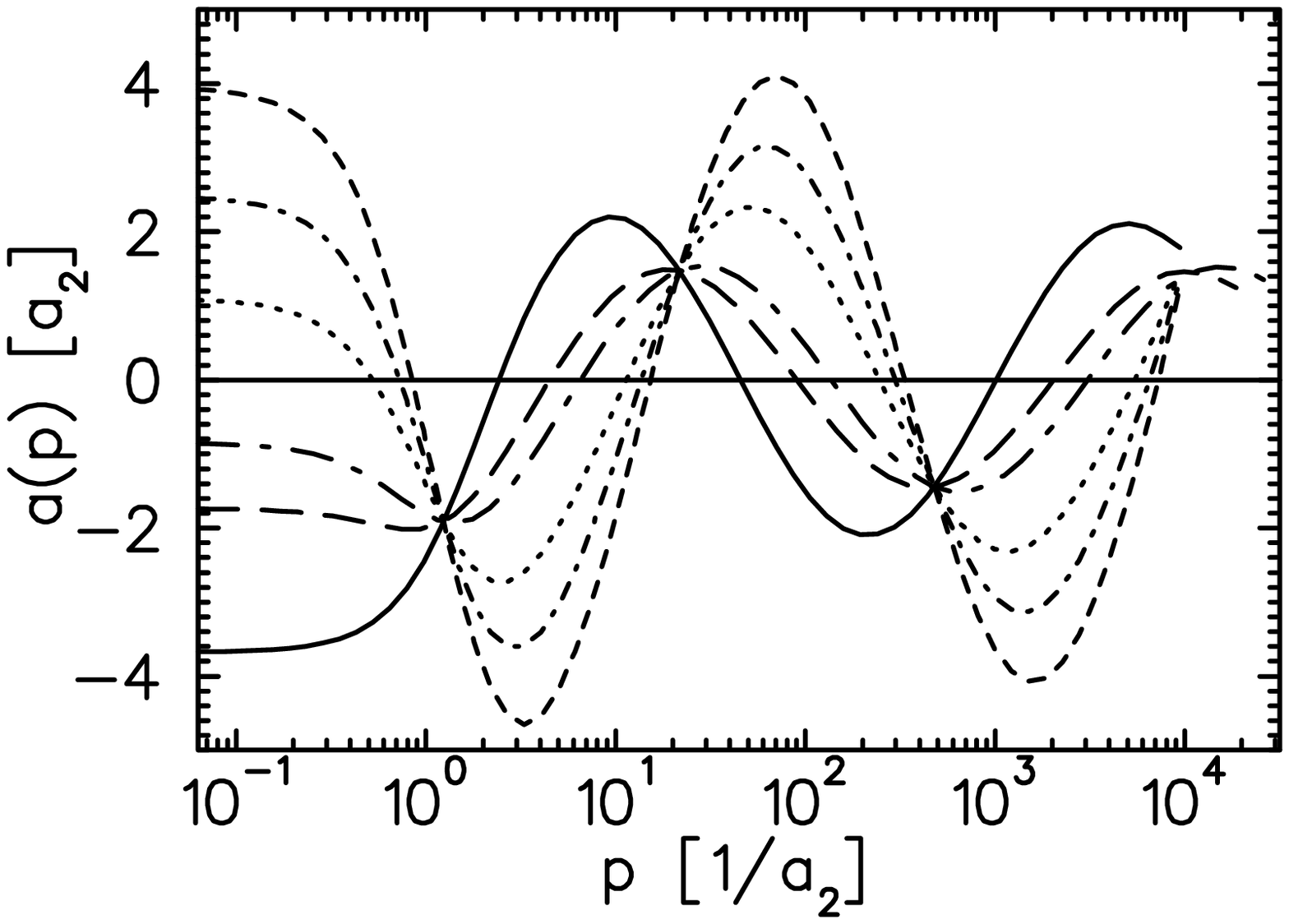}\hfill
    \includegraphics*[width=0.48\linewidth]{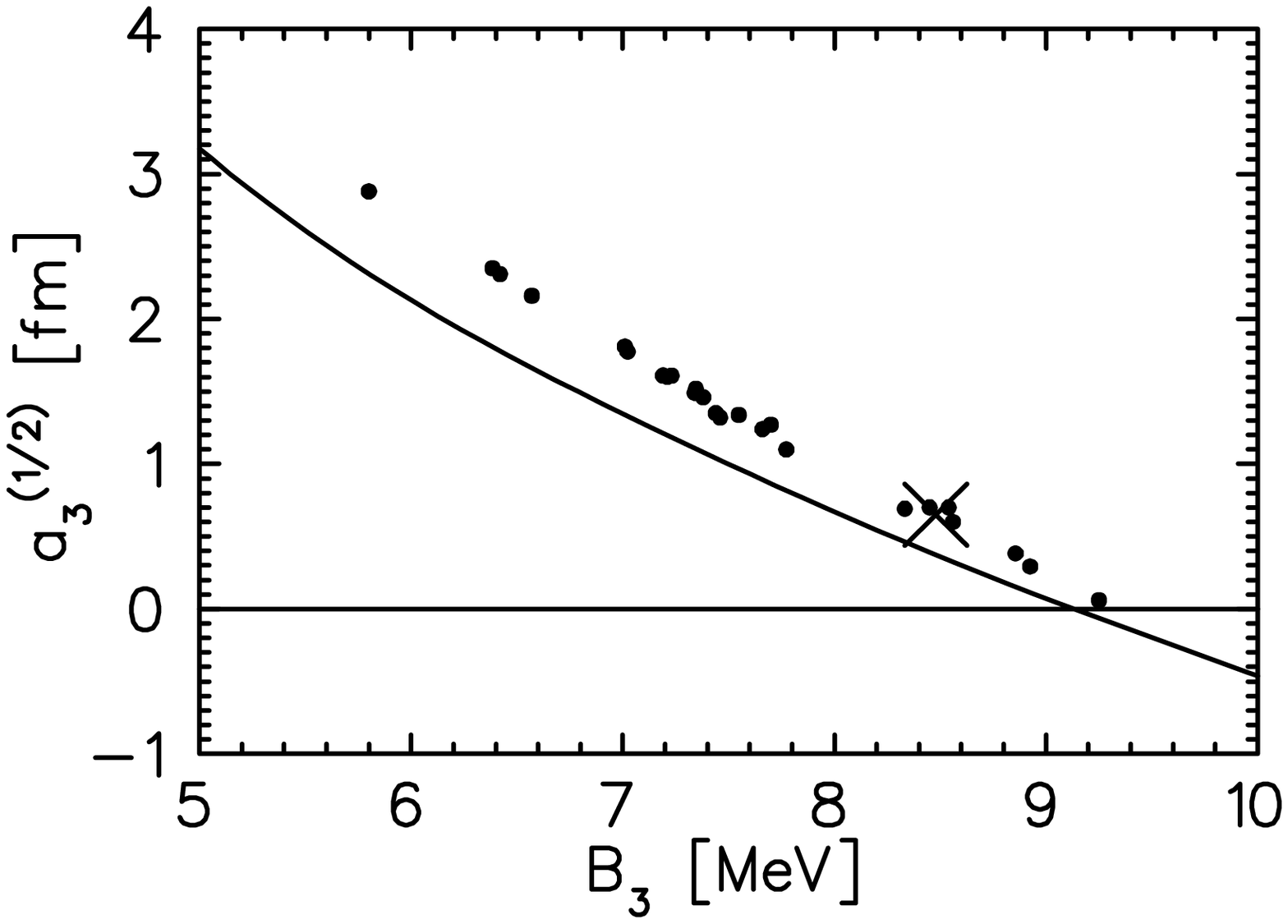}
\vspace*{-15pt}
    \caption{Left: Variation of the cut-off by a factor 3 in a numerical study
      of the na{\ia}ve off-shell amplitude Fig.~\protect\ref{fig:LOfaddeev} in
      the triton channel has a dramatic effect on the scattering length
      $a(p=0)$.  Right: Comparison of the ENT prediction for the Phillips line
      with results from various potential models. Figures from
      Ref.~\protect\cite{Stooges2}.}
    \label{fig:Phillipsline}
  \end{center}
\vspace*{-4ex}
\end{figure}

The equation is easily solved by a Mellin transformation,
$\calA_{(l,s)}(0,p)=p^{-1+s_0(l,s)}$, and indeed in most channels $s_0$ is real
so that only one solution exists which vanishes at infinite momentum and hence
is cut-off independent. However, the fact that the kernel of (\ref{UVfaddeev})
is not compact makes one suspicious whether this is always the case. Indeed,
there exists one and only one partial wave in which two linearly independent
solutions are found: the doublet $\mathrm{S}$ wave (triton) channel, where
$s_0(l=0,s=\half)=\pm1.0062\dots\ii$. Therefore in this channel, any
superposition with an arbitrary phase $\delta$ is also a solution, and we find
that
\begin{equation}
  \label{UVsolution}
  \calA_{l=0,s=\half}(k\to 0,p)\propto\frac{\cos[1.0062 \ln p+\delta]}{p}\;\;.
\end{equation}
Because each value of the phase provides a different boundary condition for
the solution of the full integral equation Fig.~\ref{fig:LOfaddeev}, the
on-shell amplitude $\calA(k,p=k)$ will also depend crucially on $\delta$.
This means that the on-shell amplitude seems sensitive to off-shell physics,
and what is more, to a phase which stems from arbitrarily high momenta. A
numerical study of the full half off-shell amplitude confirms these findings,
Fig.~\ref{fig:Phillipsline}. This cannot be.

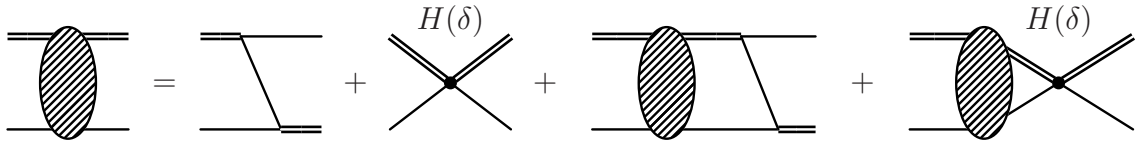
\begin{figure}[!ht]
  \begin{center}
    \vspace{-4ex}
    $\setlength{\unitlength}{0.55pt}
\feynbox{104\unitlength}{
            \begin{fmfgraph*}(104,64)
              \fmfleft{i2,i1}
              \fmfright{o2,o1}
              \fmf{double,width=thin,tension=3}{i1,v1}
              \fmf{double,width=thin,tension=1.5}{v1,v3,v2}
              \fmf{double,width=thin,tension=3}{v2,o1}
              \fmf{vanilla,width=thin}{i2,v4,o2}
              \fmffreeze
              \fmffreeze
              \fmf{ellipse,rubout=1}{v3,v4}
              \end{fmfgraph*}}
            =
            \feynbox{104\unitlength}{
            \begin{fmfgraph*}(104,64)
              \fmfleft{i2,i1}
              \fmfright{o2,o1}
              \fmf{double,width=thin,tension=4}{i1,v1,v2}
              \fmf{vanilla,width=thin}{v2,o1}
              \fmf{double,width=thin,tension=4}{v3,v4,o2}
              \fmf{vanilla,width=thin}{i2,v3}
              \fmffreeze
              \fmf{vanilla,width=thin}{v2,v3}
              \end{fmfgraph*}}
            +
            \feynbox{104\unitlength}{
            \begin{fmfgraph*}(104,64)
              \fmfleft{i2,i1}
              \fmfright{o2,o1}
              \fmf{double,width=thin}{i1,v,o1}
              \fmf{vanilla,width=thin}{i2,v,o2}
              \fmfv{decor.shape=circle,decor.filled=full,decor.size=2thick,
                label=$\blue{H(\delta)}$,label.angle=90,label.dist=0.5h}{v}
              \end{fmfgraph*}}
            +
            \feynbox{192\unitlength}{
            \begin{fmfgraph*}(192,64)
              \fmfleft{i2,i1}
              \fmfright{o2,o1}
              \fmf{double,width=thin,tension=3}{i1,v1}
              \fmf{double,width=thin,tension=1.5}{v1,v6,v5}
              \fmf{double,width=thin,tension=3}{v5,v2}
              \fmf{vanilla,width=thin}{v2,o1}
              \fmf{vanilla,width=thin}{i2,v7}
              \fmf{vanilla,width=thin,tension=0.666}{v7,v4}
              \fmf{double,width=thin,tension=4}{v4,v3,o2}
              \fmffreeze
              \fmffreeze
              \fmf{vanilla,width=thin}{v4,v2}
              \fmf{ellipse,rubout=1}{v6,v7}
              \end{fmfgraph*}}
            +
            \feynbox{192\unitlength}{
            \begin{fmfgraph*}(192,64)
              \fmfleft{i2,i1}
              \fmfright{o2,o1}
              \fmf{double,width=thin,tension=3}{i1,v1}
              \fmf{double,width=thin,tension=1.5}{v1,v6}
              \fmf{phantom,width=thin,tension=1.5}{v6,v5}
              \fmf{phantom,width=thin,tension=3}{v5,v2}
              \fmf{phantom,width=thin}{v2,o1}
              \fmf{vanilla,width=thin}{i2,v7}
              \fmf{phantom,width=thin,tension=0.666}{v7,v4}
              \fmf{phantom,width=thin,tension=4}{v4,v3,o2}
              \fmffreeze
              \fmffreeze
              \fmf{vanilla,width=thin}{v7,v,o2}
              \fmf{double,width=thin}{v6,v,o1}
              \fmfv{decor.shape=circle,decor.filled=full,decor.size=2thick,
                label=$\blue{H(\delta)}$,label.angle=90,label.dist=0.5h}{v}
              \fmf{ellipse,rubout=1}{v6,v7}
              \end{fmfgraph*}}
$
\vspace*{-15pt}
    \caption{The cut-off independent Faddeev equation in the
      triton channel.} 
    \label{fig:LOtriton}
  \end{center}
\vspace*{-4ex}
\end{figure}

Because physics must be independent of the cut-off chosen, this sensitivity of
the on-shell amplitude on UV properties of the solution to
(\ref{fig:LOfaddeev}) must be remedied by adding a counter term. And since the
power counting in the two nucleon sector is fixed, one can show that a
necessary and sufficient condition to render cut-off independent results is to
promote the three body force (\ref{threebodyterm}) to LO, $H(\delta)\sim
Q^{-2}$, and absorb all phase dependence into it, see Fig.~\ref{fig:LOtriton}.
The variation of $H(\delta)$ with the cut-off or phase is known analytically
from (\ref{UVsolution}), but its initial value is unknown. Therefore, one
physical scale $\bar{\delta}$ must be determined experimentally. This one, new
free parameter explains why potential models which provide an accurate
description of $\mathrm{NN}$ scattering can vary significantly in their
predictions of the triton binding energy $B_3$ and three body scattering
length $a_3^{(1/2)}$ in the triton channel, although all of them lie on a
curve in the $(B_3,a_3^{(1/2)})$ plane, known as the Phillips line,
Fig.~\ref{fig:Phillipsline}. Determining $H(\bar{\delta})$ by fixing the three
body scattering length to its physical value, the triton binding energy is
found in \ENTNoPion to be $8.0\;\MeV$ at LO.

\setlength{\unitlength}{1pt}
    
It must again be stressed that the three body force of strength $H(\delta)$
was added not out of phenomenological needs.  It cures the arbitrariness in
the off-shell and UV behaviour of the two body interactions which would
otherwise contaminate the on-shell amplitude.  Just as the off-shell behaviour
of the two body amplitude Fig.~\ref{fig:LOfaddeev}, the strength $H(\delta)$
is arbitrary, and only the sum of two and three body graphs is physically
meaningful. 

Summing the deuteron bubbles, each graph in the upper line of
Fig.~\ref{fig:LOfaddeev} behaves like $p^{-2}$ in the UV. But the solution of
the Faddeev equation goes like $p^{-1+s_0}$ with irrational (or even complex)
$s_0(l,s)$~\cite{hgtriton} and is hence more than the na{\ia}ve sum of graphs.

The limit cycle thus encountered in the triton channel is a new
renormalisation group phenomenon and also explains the Efimov and Thomas
effects~\cite{review,Stooges2}. However, the phenomenon occurs only in the
triton channel. In all other partial waves, three body forces enter only at
higher orders~\cite{hgtriton}.

\subsection{Partial Waves in Neutron--Deuteron Scattering}

In the three body sector, the equations to be solved in \ENTNoPion and
ENT(KSW) are computationally trivial and can furthermore be improved
systematically by higher order correction which involve only (partially
analytic, partially numerical) integrations, in contradistinction to
many-dimensional integral equations arising in other approaches. A comparative
study between the theory with explicit, perturbative pions (ENT(KSW)) and the
one with pions integrated out was performed~\cite{pbhg} in the spin quartet
$\mathrm{S}$ wave for momenta of up to $300\;\MeV$ in the centre-of-mass frame
($E_{\mathrm{cm}}\approx70\;\MeV$). As seen above, the two formulations are
identical at LO. Because three body forces enter only at high orders, this
channel is completely determined by two body properties at the first few
orders and no new, free parameters enter.

\begin{figure}[!htb] 
  \begin{center}
    \vspace{-2ex}
  \includegraphics*[width=0.52\textwidth]{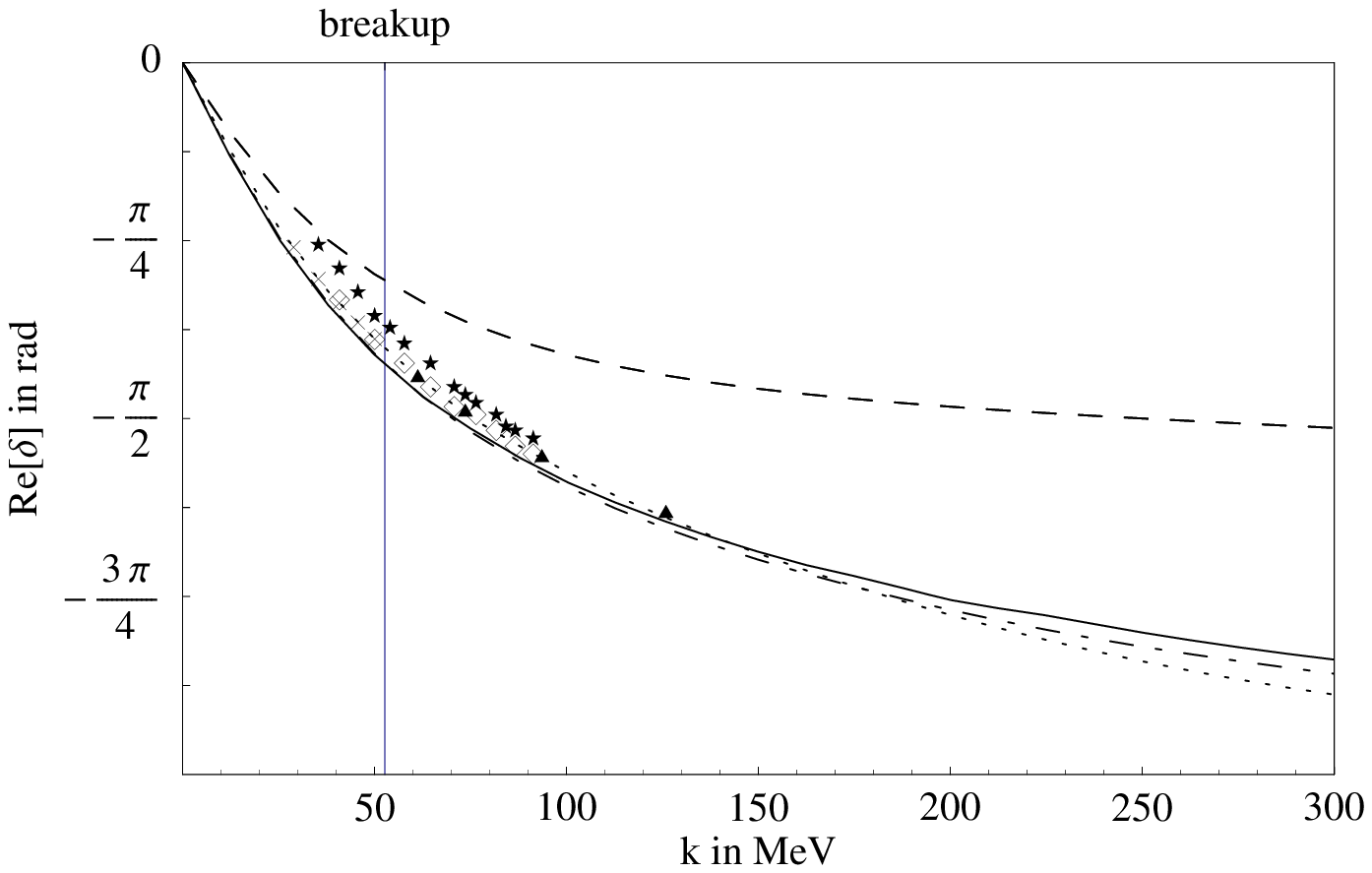}
    \hfill
    \includegraphics*[height=0.44\textwidth,angle=90]{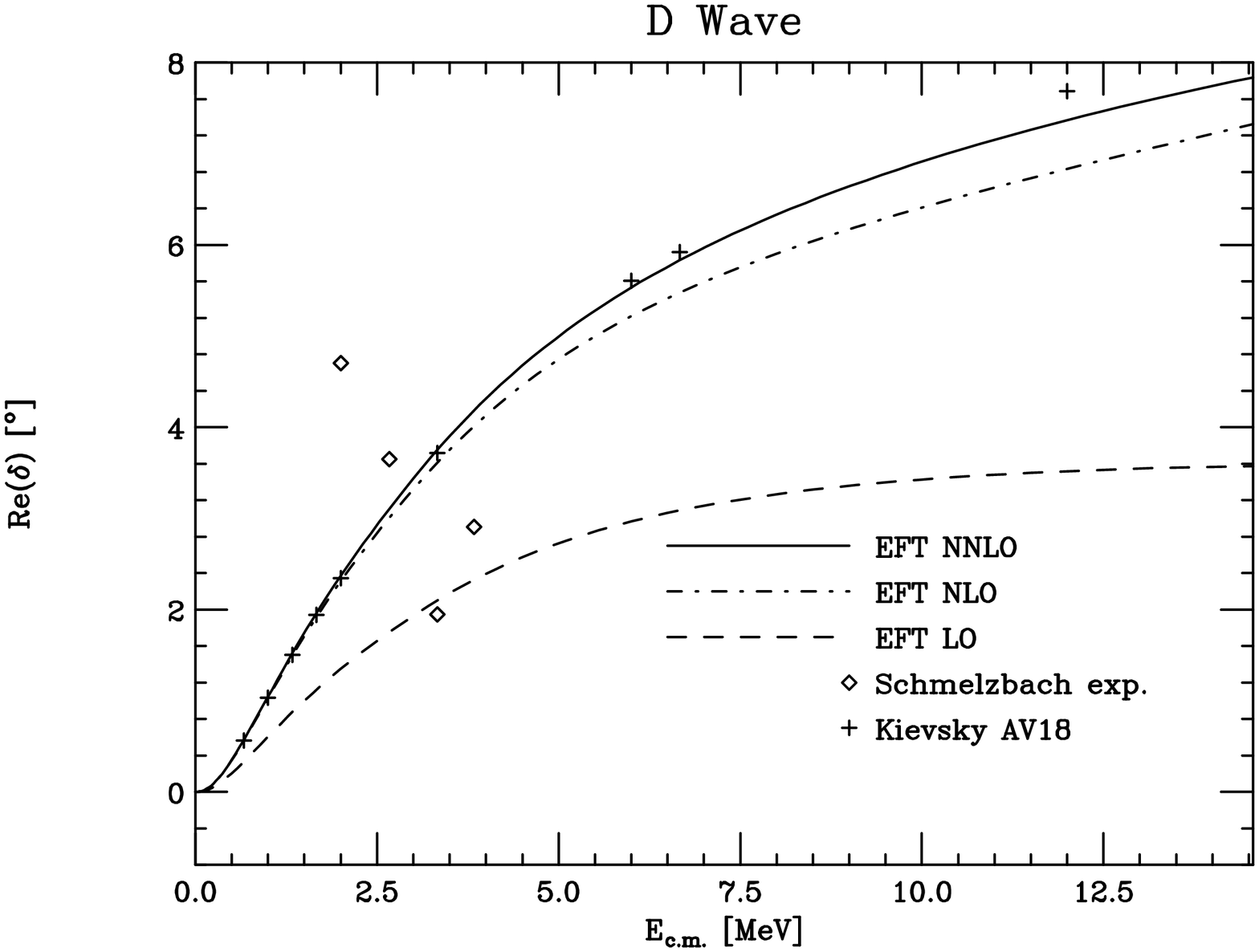}
\vspace{-15pt}
\caption{Real parts of the quartet \protect$\mathrm{S}$\protect~\cite{pbhg}
  and doublet \protect$\mathrm{D}$\protect~\cite{pbfghg} wave phase shifts in
  \protect$\mathrm{nd}$ scattering. Legend left: Dashed: LO; solid
  (dot-dashed) line: NLO with perturbative pions (pions integrated out);
  dotted: N2LO without pions. Realistic potential models: squares, crosses,
  triangles.  Stars: $\mathrm{pd}$ phase shift analysis.}
\label{fig:delta}
  \end{center}
\vspace*{-4ex}
\end{figure}

The calculation with/without explicit pions to NLO/N2LO shows convergence: For
example, the scattering length is $a({}^4S_\frac{3}{2},\mathrm{LO})=(5.1\pm
1.5)\;\fm$, and at NLO with (without) perturbative pions
$a({}^4S_\frac{3}{2},\mathrm{NLO,KSW})=(6.8\pm 0.7)\;\fm$
($a({}^4S_\frac{3}{2},\mathrm{NLO},\NoPion)=(6.7\pm 0.7)\;\fm$). At
N2LO,~\cite{Stooges2} report
$a({}^4S_\frac{3}{2},\mathrm{N2LO},\NoPion)=(6.33\pm 0.1)\;\fm$, and the
experimental value is $a({}^4S_\frac{3}{2},\mathrm{exp})=(6.35\pm 0.02)\;\fm$.
Comparing the NLO correction to the LO scattering length provides one with the
familiar error estimate at NLO: $(\frac{1}{3})^2\approx 10\%$. The NLO
calculations with and without pions lie within each other's error bar. The
N2LO calculation is inside the error ascertained to the NLO calculation and
carries itself a theoretical uncertainty of about $(\frac{1}{3})^3\approx
4\%$. The calculation of pionic corrections in ENT(KSW) shows that they are --
although formally NLO -- indeed much weaker. The difference to \ENTNoPion
should appear for momenta of the order of $\mpi$ and higher because of
non-analytic contributions of the pion cut, but those seem to be very
moderate, see Fig.~\ref{fig:delta}. This and the lack of data makes it
difficult to assess whether the KSW power counting scheme to include pions as
perturbative increases the range of validity over the pion-less theory.

\noindent
Finally, the real and imaginary parts of the higher partial waves
$l=1,\dots,4$ in the spin quartet and doublet channel were found~\cite{pbfghg}
in a parameter-free calculation, see Fig.~\ref{fig:delta}.  Within the range
of validity of this pion-less theory, convergence is good, and the results
agree with potential model calculations (as available) within the theoretical
uncertainty. That makes one optimistic about carrying out higher order
calculations of problematic spin observables like the nucleon-deuteron vector
analysing power $A_y$ where EFT will differ from potential model calculations
due to the inclusion of three-body forces.

\section{Outlook}

Many questions remain open: Does the power counting Nature chose include pions
perturbatively or non-perturbatively? At the moment, some people advocate a
mixture.  To investigate processes with pions in the initial or final state
might be helpful~\cite{bbhg}. How to extend the analysis in the triton
channel systematically to higher orders? Technically, how to regularise a
Faddeev equation numerically with external currents coupled in a field theory?
How do four body forces scale? Extend to nuclear matter!


\end{fmffile}
\end{document}